\documentclass[10pt,conference]{IEEEtran}
\makeatletter
\IEEEoverridecommandlockouts

\def\ps@headings{%
\def\@oddhead{\mbox{}\scriptsize\rightmark \hfil \thepage}%
\def\@evenhead{\scriptsize\thepage \hfil \leftmark\mbox{}}%
\def\@oddfoot{}%
\def\@evenfoot{}}
\makeatother 

\usepackage{color}
\usepackage{graphicx}               % For using graphics like figures
\usepackage{algorithm}
\usepackage{algpseudocode}
\usepackage[cmex10]{amsmath}
\usepackage{soul}

\ifCLASSINFOpdf
\else
\fi

\usepackage{algorithm}

\usepackage{algpseudocode}
\usepackage{caption}

\usepackage{url}
\begin{document}
\title{Collaborative Load Management in \\Smart Home Area Network \vspace{-0.3cm}}

\author{\IEEEauthorblockN{Jagnyashini Debadarshini}
\IEEEauthorblockA{Indian Institute of Technology Bhubaneswar\\
jd12@iitbbs.ac.in} \and \IEEEauthorblockN{Sudipta Saha}\IEEEauthorblockA{Indian Institute of Technology Bhubaneswar\\sudipta@iitbbs.ac.in
} }

\maketitle



% \fi
\IEEEpeerreviewmaketitle

\begin{abstract}
An efficient Home Area Network (HAN) acts as a base of an \textit{Advanced Metering Infrastructure} (AMI). A HAN not only facilitates AMI with efficient real-time monitoring of the electricity consumption but also manages the load profile of the whole system. However, the existing works on implementing HAN are mostly centralized and suffer from well-known problems. In this work, we propose an IoT-based efficient decentralized
strategy using synchronous transmission to practically realize HAN. An inter-device coordination strategy is proposed to minimize the peak load as well as reduce the sudden changes in the overall system without compromising the user’s requirements. Through experiments over IoT-testbeds, we demonstrate that the proposed strategy can reduce the peak load upto 50\% and reduce the load variations upto 58\% for even a high and random rate of requests for execution of power-hungry house appliances.
\end{abstract}


\section{Introduction}
\label{sec:intro}
The \textit{Advanced Metering Infrastructure} (AMI) is a very significant component of a smart electric-grid system. It enables a grid-system to closely monitor as well as manage the consumption of the electricity% in the \textit{distribution sub-system} 
  \cite{ghosal2019key}. The operation of an AMI fundamentally starts with the \textit{Home Area Network} (HAN) which binds all the electrical equipment inside a defined customer premise. An efficient HAN, thus, can not only bring a close real-time view of the consumption of the electricity, but also an appropriate management of the same which can largely avoid sudden rise or drop in the electric-load. %electricity consumption to maintain a stable frequency. 

Functionality of a HAN has been realized so far mostly in a centralized manner where all consumption information are recorded centrally. Moreover, any new request is routed to a central scheduler which subsequently takes a scheduling decision based on the current status \cite{makhadmeh2019optimization}. Various optimization methods are used for deriving such schedule \cite{molderink2009domestic,zhu2012integer} to run the appliances in the best possible manner considering peak-load, electricity pricing, users' comfort etc. Note that these proposed strategies mostly evaluated through simulation based experiments or theoretically or real-experiments considering very few number of real-devices \cite{makhadmeh2019optimization}. 

A centralised solution for HAN, although works, bears a certain known set of issues such as, single-point-of-failure, costly and in-flexible implementation etc. Apart from these, various parameters in an electrical system face dynamic changes, e.g., customer's demand, physical status of the electrical appliances, electricity pricing etc. Existing HAN systems mostly use traditional \textit{Asynchronous-Transmission} (AT) based communication protocol for exchanging messages between the devices and the central-administrator. Under this model, frequent and fast communication between the electrical appliances and the central controller becomes a significant problem which acts as a bottleneck in supporting the underlying dynamic nature of the system. As the number of devices increases, such difficulties also increase in proportion.

In this work, we propose a decentralized, flexible, cost-effective and easy-to-install strategy as an alternative way to realize HAN. In contrast to the traditional AT based approaches, in the current work we propose the use of \textit{Synchronous-Transmission} (ST) based communication \cite{zimmerling2020synchronous} for a fruitful decentralized coordination among the devices. In the following, we first provide a sketch of our proposed design followed by some evaluation results demonstrating its use in load-management through a simple inter-device coordination.

\section{Design}
We broadly categorize the electrical appliances into two types: \textit{Type-1} and \textit{Type-2}. Under Type-1, the devices require to instantly turned ON as per the request of the customers. Some of these devices may consume high power, e.g., hair-dryer, blender/mixer etc. But majority of the devices under Type-1 are usually less power hungry, e.g., fan, TV, laptop etc. Type-2 devices, in contrast, mostly consume high power but their execution can be managed/scheduled, e.g., ACs, room heaters, water heaters/coolers, fridge etc., fall in this category.

We assume that an electrical appliance gets connected with electricity outlet through a specially designed \textit{Device-Interface} (DI) \cite{pipattanasomporn2012demand}. A DI is supposed to be equipped with an 802.15.4 compatible low-power RF transceiver. The active DIs in a certain premise are supposed to form a \textit{multi-hop IoT-system}. %Figure \ref{fig:han1}(a) shows a sample scenario. 
In this work we pay special attention to the Type-2 devices for load-management. Many of these Type-2 devices internally follow a duty-cycling of their power-hungry module. For instance, to maintain a certain temperature inside a room, an AC would turn its compressor ON/OFF as per the current temperature. A room-heater or a water-heater also follow the same principle internally. We assume that the DI connected to a Type-2 device has control over this duty-cycling, i.e., a DI is able to turn the main unit of the device (e.g., the compressor of an AC or fridge) ON/OFF as per requirement.

To simplify the scenario, we introduce two constraints for a Type-2 device, namely, \textit{min-Duty-Cycle-Duration} (minDCD) and \textit{max-Duty-Cycle-Period} (maxDCP). The minDCD is the minimum time a device is supposed to stay ON once it starts while maxDCP is the maximum length of the duty-cycling period. Based on environmental factors, user demands, as well as current device status, minDCD and maxDCP may vary with time. For example, to achieve a target temperature of 20$^{\circ}$C, the maxDCP would be lesser compared to a target temperature of 30$^{\circ}$C when the external temperature is 40$^{\circ}$C. It would also vary for different devices. Considering all such dynamic situations, developing an efficient schedule in a decentralized fashion is a challenging job. 
\vspace{-0.2cm}
\begin{figure}[htbp]
\begin{center}
\includegraphics[angle=0,width=0.48\textwidth]{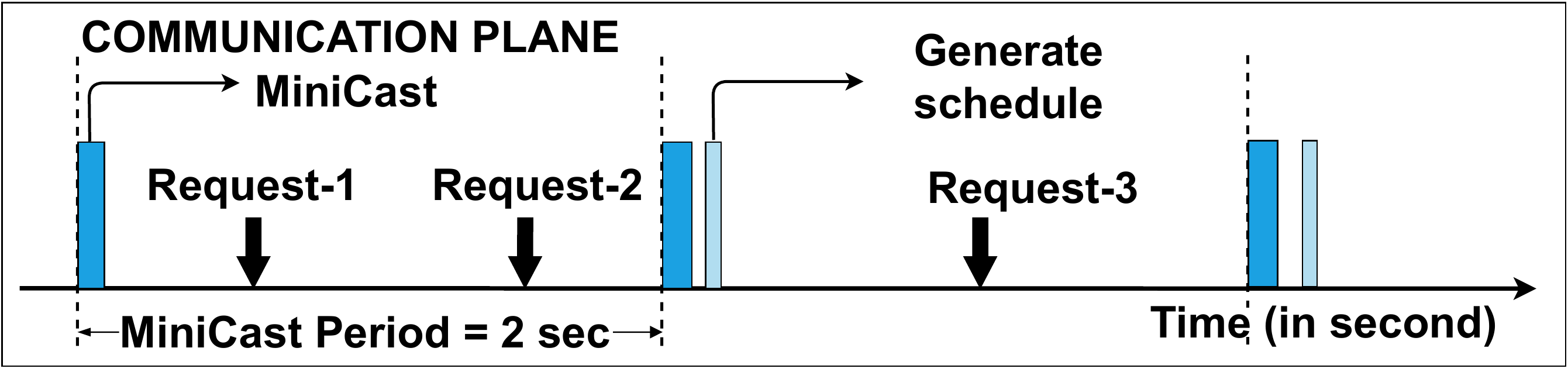}
\vspace{-0.2cm}
\end{center}
\caption{Communication Plane using MiniCast }
\label{fig:han1}
\vspace{-0.2cm}
\end{figure}

We divide the functionality of the system into two planes - \textit{Communication Plane} (CP) and \textit{Execution Plane} (EP). CP enables a very quick sharing of the status data of the devices among all the DIs as well as let all the DIs know about any new request from any user. These data are then used to form an efficient schedule that would clearly specify which device should stay ON or OFF in the next \textit{period}. In EP, the devices follow the developed schedule. Figure \ref{fig:han1} shows the CP where ST based protocol MiniCast \cite{saha2017efficient} is used to achieve all-to-all data sharing among the DIs. The protocol is repeated every 2 second. The algorithm used to build the schedule starts just after that. It is quite simple and mainly tries to coordinate the ON periods of the duty-cycles of the active devices with each other in a way that multiple-requests instead of getting stacked on each other, gets scheduled one by one. Exploiting the concept of minDCD, the algorithm also ensures that execution of at least one instance (minDCD) of each active device and newly requested device should take place within a single period of maxDCP in an organized way to avoid sudden rise in the load. The total load thus increases in small steps.

% ***This way, based on the value of the minDCD, a new request may have to wait for a little amount of time. This ensures that there is no sudden rise or fall in the load.

% We assume a homogeneous setting where all devices have same minDCD and maxDCP. Figure  \ref{fig:han1}(c) shows the \textit{Execution phase} which starts after the schedule is decided. 

% It shows the load profile while execution of two Type-2 devices with coordinated scheduling and without it. In coordinated scheduling, the load increases as a step function and decrease in that way also. This is realized by developing a schedule (OF-OFF timing for each device) in a decentralized way and following that by all the cooperating devices. 

%Thus, any new request from an user would almost instantly become available to all the devices. Whenever any new device requests comes it is shared with all the other devices along with the status for the currently running devices in the next minicast iteration. When a user requests to turn a certain Type-2 device ON, the request first gets disseminated to all the DIs through almost instantaneously through the next instance of MiniCast round. 
%ST based communication in IoT/WSN systems have gained quite a good popularity in recent systems research . 
\section{Evaluation} 
We implemented the proposed strategy in Contiki OS for TelosB devices and experimented with the same over a publicly available IoT-testbed, FlockLab \cite{trub2020flocklab} comprising of 26 IoT-nodes. We assume that all these 26 devices represent DIs that are connected to 26 different Type-2 device each consuming 1 KWh power. Users requests for each of these 26 devices are considered to be randomly arriving. The maxDCP and minDCD are assumed to be 30 mins and 15 mins respectively for all the devices. We consider three scenarios based on the rate of arrivals of the users requests, high (30 requests/hour), moderate (18 requests/hour) and low (4 requests/hour).
\begin{figure}[htbp]
\begin{center}
\includegraphics[angle=0,width=0.5\textwidth]{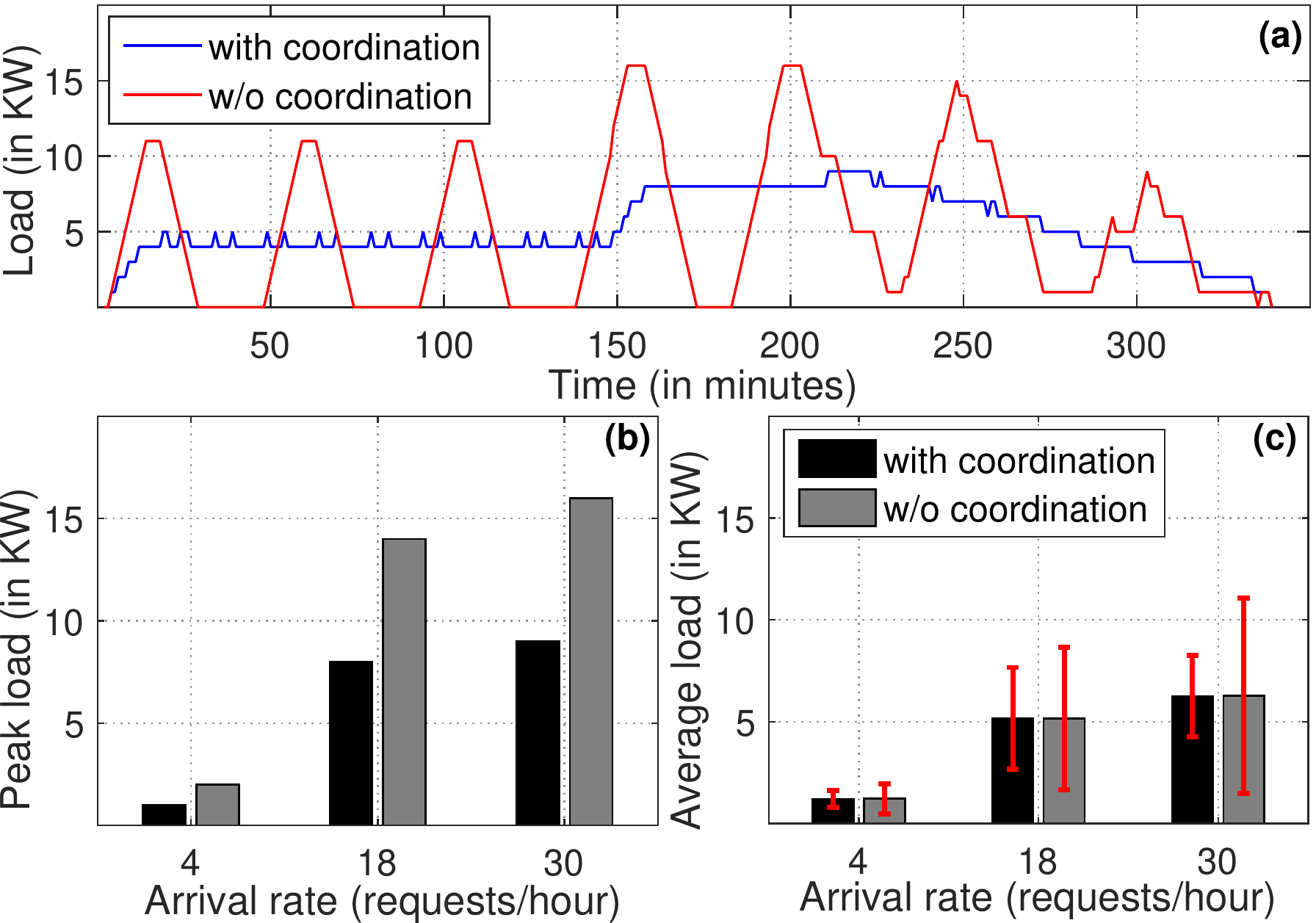}
\end{center}
\vspace{-0.3cm}
\caption{Part (a) shows the variation of load over a duration of 350 minutes with high arrival rate of jobs , part(b) shows the variation of peak load with different arrival rate of jobs, and part (c) shows the average power load and the deviation value.}
\vspace{-0.5cm}

\label{fig:plot}
\end{figure}

Figure \ref{fig:plot}(a) shows the variation of the total system-load over 350 minutes for the high-rate of arrivals of the request with coordinated duty-cycling and without the same. Figure \ref{fig:plot}(b), and \ref{fig:plot}(c) show the peak load and the average load, respectively for different request rates over the full experiment. The error bars in Figure \ref{fig:plot}(c) show the degree of variations in the load. It can be observed that proposed coordination strategy reduces the peak load upto 50\%, while keeping average load the same. It can be also seen that the standard deviation of total-load over time drastically drops upto 58\% ensuring much stable and smoother functioning of the system.

\section{conclusion}

A Home-Area-Network is in charge of tying up all the electrical appliances together to closely monitor their behavior as well as managing the system load. Centralized realization of HAN, although works, is not a preferable solution. In this work, we propose an alternative solution where the electrical devices within a certain premise are supposed to form an IoT-system. To cope up with the highly dynamic situation we  demonstrate the use of a synchronous-transmission based data sharing protocol for decentralized coordination load-management. Real testbed based evaluation shows that the proposed strategy can reduce the peak load upto 50\% and make the overall process much more smooth.

\end{document}